# Towards Dynamic Urban Scene Synthesis: The Digital Twin Descriptor Service

Ioannis Tsampras
ECE Department
University of Patras
Patras, Greece
ioannis.tsampras@ac.upatras.gr

Georgios Stergiopoulos
ECE Department
University of Patras
Patras, Greece
gstergiopoulos@ac.upatras.gr

Tanya Politi
ECE Department
University of Patras
Patras, Greece
tpoliti@upatras.gr

Spyros Denazis
ECE Department
University of Patras
Patras, Greece
sdena@upatras.gr

*Abstract*— Digital twins have been introduced as supporters to city operations, yet existing scene-descriptor formats and digital twin platforms often lack the integration, federation and adaptable connectivity that urban environments demand. Modern digital twin platforms decouple data streams and representations into separate architectural planes, fusing them only at the visualization layer and limiting potential for simulation or further processing of the combined assets. At the same time, geometry-centric file standards for digital twin description, and services built on top of them, focus primarily on explicitly declaring geometry and additional structural or photorealistic parameters, making integration with evolving context information a complicated process while limiting compatibility with newer representation methods. Additionally, multi-provider federation, critical in smart city services where multiple stakeholders may control distinct infrastructure or representation assets, is sparsely supported. Consequently, most pilots isolate context and representation, fusing them per use case with ad hoc components and custom description files or glue code, which hinders interoperability. To address these gaps, this paper proposes a novel concept "Digital Twin Descriptor Service (DTDS)" that fuses abstracted references to geometry assets and context information within a single, extensible descriptor service through NGSI-LD. The proposed DTDS can provide a dynamic and federated integration of context data, representations, and runtime synchronization across heterogeneous engines and simulators. This concept paper outlines the DTDS architectural components and description ontology that enable digital twin processes in the modern smart city.

*Keywords—smart cities, digital twin, scene descriptor, data fusion, NGSI-LD, 3D representation*

## I. Introduction

Digital twins are emerging as a critical layer in the digital infrastructure of smart cities, providing the basis for situation awareness, simulation, predictive analytics, and coordinated operations across heterogeneous urban systems. Yet, despite the rapid rise of commercial offerings and open-source frameworks, modern platforms reveal persistent gaps in how they describe, connect, represent and federate urban-scale twins. To ground our discussion, in Section II we survey six prominent platforms and tooling stacks. For each, we briefly examine how scene geometry and live context data are fused, the mechanisms for dynamic updates and support for multi-provider federation. At the same time we establish common platform components, presented in Section III. In section IV we group up recurring pain points detected in relevant approaches and articulate the driving problem behind this work.

With these observations we lead to our central contribution, the Digital Twin Descriptor Service concept, a layer that unifies abstract references to representation assets with NGSI-LD [10] based context data, under a single description service with extended declarative connectivity for updates and real time updates. As argued in Section V, we display the correspondence of our design decisions, that boil down to a reference based architecture operated through the NGSI-LD standard, with the aforementioned limitations of existing platforms. Finally, Section VI outlines the implementation of a Proof of Concept that demonstrates the viability of the proposed architecture.

TABLE I. ABBREVIATIONS AND ACRONYMS

| Acronym | Meaning |
|---|---|
| ACM | Alternative Communication Mechanism |
| ADE | Application Domain Extension |
| ADT | Azure Digital Twins |
| API | Application Programming Interface |
| AWS | Amazon Web Services |
| COLLADA | Collaborative Design Activity |
| CR | Context Reference |
| CZML | Cesium Markup Language |
| CityGML | City Geography Markup Language |
| DT | Digital Twin |
| DTDO | Digital Twin Descriptor Ontology |
| DTDS | Digital Twin Descriptor Service |
| GPS | Global Positioning System |
| HTTP | Hypertext Transfer Protocol |
| IoT | Internet of Things |
| JSON | JavaScript Object Notation |
| KML | Keyhole Markup Language |
| MQTT | Message Queuing Telemetry Transport |
| NGSI-LD | Next Generation Service Interface Linked Data |
| OGC | Open Geospatial Consortium |
| PoC | Proof of Concept |
| REST | Representational State Transfer |
| RR | Representation Reference |
| SDK | Software Development Kit |
| SUMO | Simulation of Urban Mobility |
| ThreeJS | Three JavaScript |
| TraCI | Traffic Control Interface |
| USD | Universal Scene Description |
| WebGL | Web Graphics Library |
| XR | Extended Reality |
| glTF | Graphics Library Transmission Format |
| gRPC | Google Remote Procedure Call |

## II. BACKGROUND

### A. Split Description Digital Twin Platforms

Major commercial platforms and open frameworks now offer digital twin services, but they often treat the 3D city model separately from the real-time data streams that drive the twin. This separation means that the spatial scene must be configured largely by hand and linked to data sources after the fact, which in turn limits the depth of integration between the live data and the 3D environment.

AWS IoT TwinMaker [1] introduces a scene composer where engineers import asset files and manually place "tags" that bind each mesh to a live entity stream (with support ranging from simple AWS native IoT data sources up to custom event connectors or even live video streaming). Telemetry then appears as labels or badges, but object motion or geometry changes must be scripted ad hoc because the 3D scene description itself is never updated by the service but lies in a custom file hidden from the user in contrast to the digital twin data which are programmatically updatable by "Component Connectors" and queryable through a knowledge graph based service. So the TwinMaker microservice supports data-rich dashboards while geometry and data remain loosely coupled.

Azure Digital Twins and the 3D Scenes Studio tool follow an almost identical pattern, ADT holds the semantic graph in "Digital Twins Definition Language", whereas 3D Scenes Studio stores a separate scene package, "a 3D file plus a configuration file", in Azure Blob Storage [2]. Elements inside the scene are then linked to twin properties so behaviors (color rules, widgets) are evaluated on the client side, meaning the spatial model stays static and must be updated manually (e.g., if objects move).

OpenTwins [3] is the most complete open-source alternative, built on Eclipse Ditto Digital Twin Information Broker, yet its 3D layer is delivered as a Unity WebGL scene wrapped in a custom dashboard panel plugin, where developers simply name each Unity mesh with a sensor serial number, causing geometry to live in a Unity asset, curated separately from the Ditto thing graph, again limiting out-of-the-box kinematic behavior, native scene synthesis or programmatic access to the scene.

### B. Fused Description Digital Twin Platforms

Across all three platforms mentioned above the pattern is the same: data first, scene second, joined only by manual bindings, an architecture perfectly adequate for visual status boards but unsuitable for city-scale simulations or complex operations such as simulation. Outside this pattern, two "true" digital twin ecosystems are noteworthy:

- The CityGML Urban digital twin definition format [5] and related tools and services [9].
- The open-USD scene descriptor [4] and the NVIDIA Omniverse ecosystem [8].

CityGML has long provided a rich, semantically-typed schema for urban digital twins describing both hierarchical geometries and context attributes. Version 3.0 [5] introduced the dynamizer Application Domain Extension so that any attribute can be overridden by a time-series or a live sensor link. Yet, even proponents acknowledge that the mechanism is verbose [11] and that support is still thin. Dynamic data ends up in external services while the file descriptor remains static. When the model is loaded into a purpose made database to serve CityGML objects, extra tables are created for Dynamizer payloads through custom extensions of the database [12][13] and clients typically poll the database to obtain updates. In practice, most projects sidestep the Dynamizer ADA by managing sensor data in parallel and then fusing it with the city model at the application level. This is more evident when looking at the most common delivery methods of CityGML assets, in Digital Twin pilots, through tiled formats like "… 3D Tiles or a tiled collection of KML, COLLADA or glTF files …" that hold only the digital representation information, ignoring the fields that would support dynamic updates [14]. Additionally, the updating of geometric aspects that include changes to the geometries themselves or positioning information has been supposed to require new iterations of the entire file through a versioning system [15] while specifically for moving objects, the compatibility with the Dynamizer ADE has been quite limited [14].

On the other hand, open-USD [4] is a layered, scene-graph file format that encodes 3D geometry, materials, cameras, lights and arbitrary metadata. Its composability and non-destructive layering aspects have established it as an attractive format for collaborative scene production pipelines. The Nvidia Omniverse ecosystem [8] is employing open-USD and utilizes the arbitrary metadata fields in assets to include application specific parameters and data, fusing context information and 3D representations in one state served to through its core "Nucleus" component and its "Live-Sync" protocol. However, correspondence of asset attributes and context information is not standardized and heavily rely on per-application interpretation. Furthermore, the attributes are file based, and updates are treated as versioning changes limiting throughput capabilities for large scale connected digital twins. Repository-centric access control makes clients mount specific Nucleus paths instead of querying a catalog and multi-tenant city-scale twins must revert to project folders or separate Nucleus instances, fracturing the global scene graph and confining the ecosystem to design collaboration and small-scale visualization.

A further limitation of both open-USD and CityGML is the geometry-as-explicit mesh mindset. The files must enumerate geometric primitives (vertices, triangles, etc) meaning that compatibility for representation methods is strict and novel techniques or newer formats such as 3D Gaussian Splats [16] are not supported (although experimental support has recently been announced in open-USD) without ecosystem-wide schema updates. Similarly, mechanisms for multiple versions or providers for a representation asset are not present. These inherit disadvantages of including the geometry definition directly inside the digital twin descriptor, that additionally require first-hand ownership or access to the asset, reflect an architectural choice that contrasts the modern urban digital environment where geospatial assets are commonly sourced at run-time as a service through web APIs from a range of providers and oftentimes directly by the client.

Finally, a slightly different, approach is presented by the Cesium ecosystem. While a fused digital twin service is not provided, the tools to describe and host such functionality are present. Representation assets can be hosted through Cesium Ion [7] and scene information can be described through the CZML JSON format [17], mainly including positioning information but the format can be extended to include context. It is meant to be sufficiently client-agnostic that compatible clients can render the scene described therein but it is mainly utilized through the Cesium web client. Live updates to the twin are also possible through lightweight incremental streaming. While the format has not been yet standardized the overall ecosystem adoption in modern urban digital twin visualization is widespread due to its simplicity, variety, multimodal support, and plugins for popular engines. However, while the representation asset hosting service, Cesium Ion, is well established as a product, the CZML streaming and state management is left to the developer, meaning that there is no complete component to store and update the state of the twin.

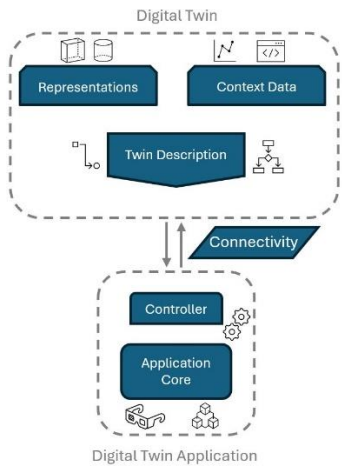

Fig. 1. The Six Digital Twin Platform Key Concepts

### III. SIX KEY CONCEPTS

We observe that across all digital twin platform implementations a pattern emerges. We can distinguish six key concepts that take form in each case reviewed.

1. Representation Assets
2. Context Data
3. Twin description
   (Positioning, hierarchy, metadata)
4. Connectivity
   (interfaces and protocols exposed to the client)
5. Application Core
   (e.g., rendering engine, simulation software etc.)
6. Application Controller
   (often named "connector", "extension", "plugin")

These concepts may appear convoluted together in one or more architectural components. For example, in the Omniverse ecosystem, both (1),(2) and (3) are coupled together in a USD file. In contrast, where the TwinMaker is concerned, the twin description (3) is a simple JSON file, the Representations (1) are stored as glTF binaries and Context Data (2) are retrieved through a Cloud Native connector stack. However, the six key concepts are present in all cases reviewed. This consistency signifies their importance in designing a platform capable of modern digital twin operations.

### IV. PROBLEM STATEMENT

The gap for a digital twin descriptor and corresponding service that caters to the needs of the modern urban digital landscape is apparent. Limitations of existing platforms and ecosystems include:

a) Split scene and data architectures
b) No native multitenancy and federation support
c) Restrictive explicit geometry descriptions
d) Weak support for dynamic updates
e) Runtime, licensing or hardware dependencies
f) Lack of concrete context information management
g) Limited scaling due to file-based services

A platform that lives up to the expectations of the modern urban digital twin would need to overcome these challenges and remain simple in definition and adoption process. It should be able to describe complete scenes and fuse the representation and context information in its internal state, remain free of runtime or hardware dependencies and land on top of a well-established, standardized, multitenancy capable and federated working base. Functionally, the framework should be service based utilizing modern scalable interfaces and internal state management while providing real-time asynchronous update buses to the clients in addition to a simple consumer interface. Additionally, the description should be able to link representation assets of different modalities and sizes from multiple providers and avoid defining geometries inside the digital twin description by abstracting the representation asset to a web-based resource. The client controllers should be expected to be able to run a minimal set of application specific primitive operations to connect to the service, select scenes, parse the descriptors, load supported versions of assets and set up synchronization channels in the environment of the specific digital twin application.

### V. PROPOSED CONCEPT

We envision the DTDS as an architecture, shown in Fig. 3, that assumes existing instantiations of context information and digital twin representation assets, fusing the two in a Digital Twin Descriptor and allowing versatile interaction with digital twin applications such as viewers, simulators, predictors or other smart city digital twin clients and services. The key innovation is the reference-based nature of the proposed ontology that enables digital twins to fuse context data and representation assets owned and served by different stakeholders while also providing flexibility with respect to the asset exchange format, addressing limitations (a) to (c) from section IV. Additionally, the service derived from the aforementioned ontology can handle dynamic updates in a native and established manner through the NGSI-LD standard, maintaining support for existing NGSI-LD based supporting services such as IoT Agents, versioning, persistence and analytics while benefitting from existing scaling solutions, confronting issues (d) to (g) from the previous section.

## A. Twin Descriptor

The Digital Twin Descriptor Ontology (DTDO), as presented in Fig. 2, fuses context information and representation assets in a scene-graph type structure. A scene represents a combination of assets in a specific geographic area. For a specific area there might be different scenes depending on the intended use. Meanwhile, one asset may be referenced in multiple overlapping scenes.

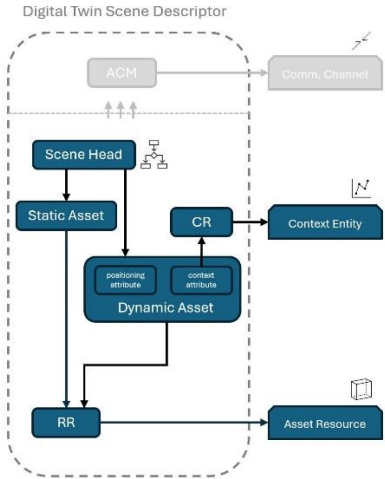

Fig. 2. Digital Twin Descriptor Ontology

The ontology starts with the "Scene Head Entity" as the root of each scene-graph. Static and dynamic asset entities are direct children and can correspond to environmental ("background") representations (e.g., streets, buildings, squares etc.) or dynamic ("foreground") objects (e.g., vehicles, user avatars, devices etc.) accordingly. The distinction between the two types of assets is a well-established design pattern [18][19] reflecting separate optimization techniques and an intrinsic differentiation in both asset generation pipelines and applicable utilization scenarios.

Both types of asset entities might include references to parent or children of the same type. Furthermore, references to one or more possible representation resources can be achieved by both asset types through the "Representation Reference Entity" (RR). Time dependent (contextual) attributes of dynamic assets, except movement and positioning information which are first order properties to sidestep complications highlighted in previous work [14], can be linked to traditional context entities through a "Context Reference Entity" (CR), avoiding data duplication. Finally, both context reference entities and positional attributes may point to one or more "Alternative Communication Mechanism Entities" (ACM), specifying additional synchronization channels besides the default NGSI-LD based methods, catering to real time dependent use cases, restrictive application runtimes and asynchronous updating functionalities as outlined in subsection E. CRs may include additional metadata regarding the source entity besides the appropriate ACMs.

## B. Context Information

To make adoption straightforward and integrate with existing work in the space we assume that context information will be either available directly as context sources to the DTDS or registered to existing context brokers to which the DTDS may reference back to (through Context Source Registrations). This choice enables compatibility with a variety of existing tools and services while even information sources that are not native to the NGSI-LD ecosystem may be integrated through appropriate connectors.

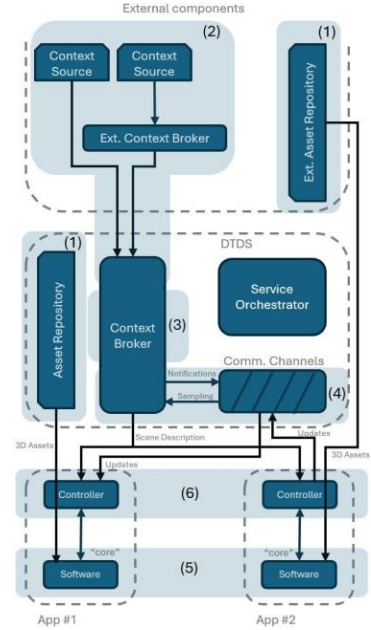

Fig. 3. Example Architecture enabled by the DTDS with the six key concepts from Section IV highlighted

## C. Asset Repositories

Furthermore, we expect digital twin representation assets to be available as addressable web resources through existing platforms including Cesium Ion [7], standardized APIs such as OGC 3D GeoVolumes [20] (currently in draft status) or purpose made asset repositories. However, it is evident that, in contrast to context information, concrete standardization of geospatial content delivery is yet to be, even for established asset exchange formats, therefore, the relevant part of the digital twin descriptors that reference representation resources should additionally describe the appropriate "Access Methods" for each resource in an "Asset Repository". These can be as simple as descriptions of RESTful API calls for asset hosting services or as complex as SDK configurations for specific ecosystems. We additionally propose the inclusion of a rudimentary model hosting service as an internal asset repository component in the DTDS to decrease the "time-to-triangle" of the platform and pave the way to a digital twin oriented standardized interface for format-agnostic representation asset hosting and delivery.

## D. Digital Twin Applications

Each digital twin application is a two-part component based around the existing implementation of the core application software and the DTDS compatible controller module, as seen in Fig. 3. The possible environments in which the core might operate (e.g., in-browser, XR device runtime, compute-based service, desktop application etc.), the runtime environment offered to the controller (e.g., separate parent or child process, software extension, proprietary scripting environment etc.), the supported representation formats and the core API accessible to the controller module are wildly varying options that need not to be defined, since, for the most part, the per-core controller modules will assume the interfacing effort. The supported representation asset types, formats, access methods and sizes, available communication mechanisms and compatible context information modalities may also differ between controllers (even for the same core application) based on the scope of the digital twin application, in addition to the specifications of the runtime environment.

*E. Communication Channels*

A common limitation of prominent environments lies in the communication controllers ability to communicate with external services due to common runtime restrictions. The DTDS could extend the NGSI-LD defined context retrieval methods to incorporate more communication mechanisms for better real time synchronization and or asynchronous updates to the clients in more restrictive runtimes and more real-time dependent use cases than those originally envisioned by the NGSI-LD standard, an effort already in place by prominent ecosystems such as FIWARE and the Orion Context Broker [21]. In this work, the ACM entity is used to define MQTT based synchronization but the corresponding parts of the ontology are grayed out to indicate that alternative communication methods should be handled by the service indirectly and not be part of the digital twin description.

## VI. PROOF OF CONCEPT

To verify that the Digital Twin Descriptor Service can mediate live cyber-physical interaction, we implemented a proof-of-concept (PoC) that fuses a copy of the physical world, a GPS tracked car, with a fleet of virtual vehicles, generated by SUMO traffic simulator [22]. This PoC demonstrates that a single DTDS instance can lead to a DT capable of orchestrating context data, real time simulation, representations and rendering.

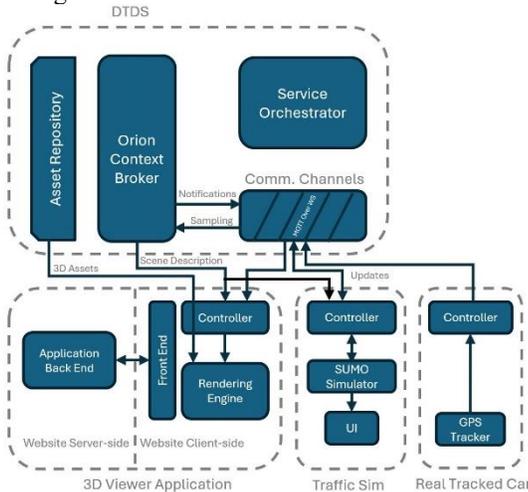

Fig. 4. Implemented PoC Architecture

In our proof-of-concept deployment Orion Context broker holds the scene graph encoded with the proposed DTDO. The controller of the GPS tracked car, through a communication channel and sampling, updates the broker for its position. SUMO's controller consumes these real-car updates and advances the simulation. Through the communication layer of the DTDS, Orion records the data published by SUMO's controller, as updates to the corresponding Dynamic Asset entity. Finally, the web based DT viewer, having configured the registered ACM, updates the rendered scene in real-time, giving the opportunity to the system operator to monitor the live interaction between physical and simulated traffic, within a single coherent view.

A real-world car, equipped with a GPS tracker streams its GPS position alongside information about its orientation (roll, pitch, yaw) over an MQTT broker. The sampling mechanism updates the corresponding Dynamic Asset entity in the service's context broker.

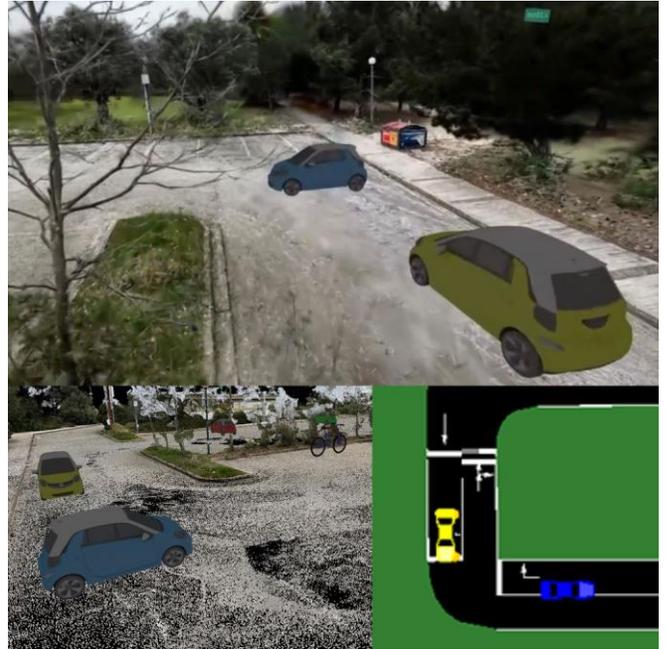

Fig. 5. The PoC Scene Visualized with a 3D Gaussian Splat layer (top), a Pointcloud layer (left) and through the SUMO GUI (right)

To utilize SUMO in a real-world scene, it's mandatory to convert the road network of the desired location, in a SUMO compatible format. Thanks to open resources and SUMO's conversion functionality this procedure can be easily performed. As a result of this, SUMO has now a translation mechanism between its own positioning system and global coordinates.

Through SUMO's TraCI Python API, a DTDS controller subscribes to the tracked car asset entity and injects the live vehicle position into the simulator. In parallel, virtual cars are generated by SUMO, the position and context information of which are posted to the DTDS as well.

The simulator, monitoring the state of both virtual and real cars, can now predict conflicts, such as collisions, and emit changes to the virtual car speed, to avoid them. Thus, a real-time interaction between the physical and virtual world is demonstrated.

For the PoC, the DT is rendered on a web-based 3D viewer compliant with the DTDS. Following the logic of the Scene Descriptor Ontology, a controller for the WebGL based

ThreeJS graphical environment was made to translate the scene descriptor into a fully functional Digital Twin scene. This implementation verifies that DTDO includes the mandatory information needed for a DT to function, display real-time data, and project simulation results.

## VII. CHALLENGES

Since interactions occur via independent HTTP requests or subscribe/notify cycles, with no transactional or ordering mechanism, clients cannot rely on a consistent, synchronized view of the context data. Additionally, NGSI-LD is strictly RESTful over HTTP, forcing workarounds for server-push updates. While Orion does offer MQTT notifications, these are only an ad-hoc unstandardized extension. Furthermore, none of the major brokers are architected for the continuous context streams required for immersive and reliable digital twins. External benchmarks of Orion under blocking update loads show round-trip times in the order of seconds with MQTT notification paths introducing additional delays [23], far too high for the sub-100ms, responsiveness expected in modern digital-twin scenarios. An extension of the NGSI-LD interface could accommodate aforementioned limitations. Finally, the question of integration with existing ecosystems could be answered through the IoT-Agent framework [10] for context data synchronization amongst platforms while representation resource interoperability can be achieved by simple resource identifiers linking assets from heterogenous platforms.

## VIII. CONCLUSIONS

This paper proposed the Digital Twin Descriptor Service as a lightweight service layer that unifies geometric asset references and NGSI-LD context into a single, extensible descriptor service. By separating heavy geometry storage from scene orchestration, while exposing a coherent scene graph, DTDS eliminates seven long-standing pain points in current urban digital twin platforms, from rigid mesh pipelines to file-bound scalability limits. A proof-of-concept implementation showcased dynamic representation assets, cyber-physical interactions and lightweight web-based visualization. Future work could improve the alternative communication methods of the DTDS, formalize the asset repository interface, crystallize schemas and introduce transactional synchronization.

## ACKNOWLEDGMENT

We would like to acknowledge that this work is funded by the European Commission, Horizon Europe, Grant Agreement No 101087257, METACITIES Excellence Hub and it was supported in the framework of the National Recovery and Resilience Plan Greece 2.0, funded by the European Union – NextGenerationEU, grant (Project: Smart Cities, TAEDR- 0536642).